\documentclass[%
  reprint,
  showpacs,preprintnumbers,
  amsmath,amssymb,
  aps,
prl,
]{revtex4-1}

\usepackage{graphicx}
\usepackage{dcolumn}
\usepackage{bm}
\usepackage{hyperref}
\usepackage{color}
\usepackage{ulem}

\usepackage{times}
\usepackage{booktabs}

\begin{document}

\preprint{APS/123-QED}

\title{
Electric Toroidal Quadrupoles in Spin-Orbit Coupled Metal Cd$_2$Re$_2$O$_7$ 
}

\author{Satoru Hayami$^1$, Yuki Yanagi$^2$, Hiroaki Kusunose$^2$, and Yukitoshi Motome$^3$}
\affiliation{
 $^1$Faculty of Science, Hokkaido University, Sapporo 060-0810, Japan \\
 $^2$Department of Physics, Meiji University, Kawasaki 214-8571, Japan \\
 $^3$Department of Applied Physics, University of Tokyo, Tokyo 113-8656, Japan
}
 
\begin{abstract}
We report our theoretical results on the order parameters for the pyrochlore metal Cd$_2$Re$_2$O$_7$, which undergoes enigmatic phase transitions with inversion symmetry breaking. 
By carefully examining active electronic degrees of freedom based on the lattice symmetry, we propose that two parity-breaking phases at ambient pressure are described by unconventional multipoles, {\it electric toroidal quadrupoles} 
(ETQs) with different components, $x^2-y^2$ and $3z^2-r^2$, in the pyrochlore tetrahedral unit. 
We elucidate that the ETQs are activated by bond or spin-current order on Re-Re bonds. 
Our ETQ scenario provides a key to reconcile the experimental contradictions, by measuring ETQ specific phenomena, 
such as peculiar spin splittings in the electronic band structure, magneto-current effect, and nonreciprocal transport under a magnetic field. 
\end{abstract}
\maketitle

{\it Introduction.---}
The spin-orbit coupling in crystals with the lack of spatial inversion symmetry, dubbed the antisymmetric spin-orbit coupling (ASOC), has attracted great interest in condensed matter physics. 
It is a source of intriguing phenomena, such as Dirac electrons at the surface of topological insulators~\cite{Hasan_RevModPhys.82.3045,Qi_RevModPhys.83.1057}, the spin Hall effect~\cite{hirsch1999spin,Sinova_PhysRevLett.92.126603}, multiferroics~\cite{Fiebig0022-3727-38-8-R01,cheong2007multiferroics,KhomskiiPhysics.2.20}, and noncentrosymmetric superconductivity~\cite{Bauer_Sigrist201201}. 
Such ASOC-related physics has been found in a variety of materials irrespective of insulators (semiconductors)~\cite{Dresselhaus_Dresselhaus_Jorio,ishizaka2011giant,furukawa2017observation,ideue2017bulk} and metals~\cite{Bauer_PhysRevLett.92.027003,Ali_PhysRevLett.109.227201,witczak2014correlated,saito2018evidence} in $p$-, $d$-, and $f$-electron systems. 
Thus, the ASOC is highly expected to bring a new route toward applications to next-generation electronics and spintronics devices~\cite{Zutic_RevModPhys.76.323, Baltz_RevModPhys.90.015005}. 

Of special interest is to control the ASOC by spontaneous inversion symmetry breaking in electronic degrees of freedom. 
Such parity breaking can generate odd-parity multipoles, e.g., magnetic quadrupoles (MQs) and electric octupoles (EOs)~\cite{Yanase_JPSJ.83.014703,hitomi2014electric,hitomi2016electric,kimura2016magnetodielectric,Kato_PhysRevLett.118.107601,Khanh_PhysRevB.96.094434,Yanagi_PhysRevB.97.020404,Hayami_PhysRevB.97.024414}. 
They provide a fertile ground for exploring new types of multipole orders~\cite{fradkin2010nematic,Hayami_PhysRevB.90.024432,Hayami_PhysRevB.90.081115,Fu_PhysRevLett.115.026401,Norman_PhysRevB.92.075113,hayami2016emergent,Watanabe_PhysRevB.96.064432,Matteo_PhysRevB.96.115156} and unconventional superconductivities~\cite{Kozii_PhysRevLett.115.207002,Wang_PhysRevB.93.134512,Wu_PhysRevB.96.144504,Sumita_PhysRevLett.119.027001}.
The pyrochlore oxide Cd$_2$Re$_2$O$_7$ is a prototype compound for such spontaneous inversion symmetry breaking in the presence of the strong spin-orbit coupling~\cite{hiroi2017pyrochlore, Fu_PhysRevLett.115.026401}. 
The system exhibits a surprisingly complex phase diagram while changing temperature and pressure, including a collection of spontaneously parity-breaking phases~\cite{Kobayashi2011super,Barisic_PhysRevB.67.245112,Yamaura_PhysRevB.95.020102,sergienko2003structural}. 
In addition, among many pyrochlores, it is the only superconductor thus far~\cite{Hanawa_PhysRevLett.87.187001,sakai2001superconductivity,Jin_PhysRevB.64.180503,hiroi2002high,hiroi2002superconducting}. 
The superconducting state also shows unconventional behavior under pressure, presumably due to the spontaneous parity breaking~\cite{Kobayashi2011super,Kozii_PhysRevLett.115.207002,Wang_PhysRevB.93.134512}.

At ambient pressure, Cd$_2$Re$_2$O$_7$ undergoes a continuous structural phase transition at $T_{\rm s 1} \sim 200$~K, from the centrosymmetric cubic phase with Fd$\bar{3}$m symmetry (phase I) to the noncentrosymmetric tetragonal one (phase II). 
As the tetragonal lattice distortion is very small, which was evaluated at most 0.05\%~\cite{Castellan_PhysRevB.66.134528}, the transition is considered to be of electronic origin. 
However, the space-group symmetry in the phase II is still controversial; it was identified as I$\bar{4}$m2 by the single-crystal X-ray diffraction (XRD)~\cite{yamaura2002low,Castellan_PhysRevB.66.134528}, powder neutron diffraction~\cite{weller2004pyrochlore}, convergent electron diffraction (CED)~\cite{Tsuda}, Raman spectroscopy~\cite{Kendziora_PhysRevLett.95.125503}, nonlinear optics~\cite{petersen2006nonlinear}, and polarizing microscope image (PMI)~\cite{matsubayashi2018formation}, while the recent nonlinear optical measurements indicated further symmetry reduction to I$\bar{4}$, I$\bar{4}$m$'$2$'$, or I$\bar{4}$m$'$2~\cite{harter2017parity,Harter_PhysRevLett.120.047601,Matteo_PhysRevB.96.115156}.
Moreover, another structural transition to the phase III occurring at $T_{\rm s 2}\sim 120$ K is also controversial; the single-crystal XRD~\cite{yamaura2002low,razavi2015effect}, CED~\cite{Tsuda}, and PMI~\cite{matsubayashi2018formation} measurements indicated a first-order transition to I$4_1$22, while the nonlinear optical measurements indicated the absence of the phase transition~\cite{petersen2006nonlinear}. 
Toward comprehensive understanding of the rich physics by spontaneous parity breaking and emergent ASOC in this compound, it is desired to resolve the experimental contradictions and clarify the origin of the enigmatic phase transitions.

In this Letter, we investigate what types of electronic instability can occur in the spin-orbit coupled metal Cd$_2$Re$_2$O$_7$ at ambient pressure from the viewpoint of odd-parity multipoles. 
Relying on the lattice symmetry by the single-crystal XRD~\cite{yamaura2002low}, we here concentrate on odd-parity multipoles with E$_u$ symmetry~\footnote{The analysis is straightforwardly applicable to another symmetry like ${\rm T}_{2u}$ suggested in Refs.~\cite{harter2017parity,Harter_PhysRevLett.120.047601}.}. 
We find that the order parameters in the phases II and III are described by electric toroidal quadrupoles (ETQs) in the tetrahedral unit of the pyrochlore structure with different components of $x^2-y^2$ and $3z^2-r^2$, respectively. 
We show that spontaneous bond or spin-current ordering on Re-Re bonds is essential to induce the ETQs. 
We also present how to detect the ETQs in experiments by elucidating ETQ-driven phenomena, such as the spin-split Fermi surface, magneto-current (MC) effect, and nonreciprocal transport (NRT) in an applied magnetic field.
 
{\it Symmetry argument.---}
First, we discuss the candidates of order parameters for the phases II and III in Cd$_2$Re$_2$O$_7$ from a symmetry point of view. 
In order to describe the electronic degrees of freedom in crystals, we introduce four types of multipoles: conventional electric (E) and magnetic (M) multipoles (polar and axial tensor, respectively), and unconventional electric toroidal (ET) and magnetic toroidal (MT) multipoles (axial and polar tensor, respectively). 
They have different parity for spatial inversion ($\mathcal{P}$) and time-reversal ($\mathcal{T}$) operations; with respect to $(\mathcal{P}, \mathcal{T})$, E multipole $Q_{lm}$ has the parity $[(-1)^l, +1]$, M multipole $M_{lm}$ has $[(-1)^{l+1}, -1]$, ET multipole $G_{lm}$ has $[(-1)^{l+1}, +1]$, and MT multipole $T_{lm}$ has $[(-1)^l, -1]$, where $l$ and $m$ are the orbital and magnetic quantum numbers, respectively ($-l\leq m\leq l$)~\cite{hayami2018microscopic,Hayami_PhysRevB.98.165110}. 
Hence, the types of symmetry breakings are systematically characterized by four types of multipoles with the different rank $l$. 
Hereafter, we use the notations for monopole ($l=0$), dipole ($l=1$), quadrupole ($l=2$), and octupole ($l=3$) as $X_0$, $(X_x, X_y, X_z)$, $(X_{u}, X_{v}, X_{yz}, X_{zx}, X_{xy})$, and $(X_{xyz}, X_x^\alpha, X_y^\alpha, X_z^\alpha, X_x^\beta, X_y^\beta, X_z^\beta)$ for E, M, ET, and MT, respectively, where $X=Q$, $M$, $G$, and $T$, and $u=3z^2-r^2$ and $v=x^2-y^2$~\cite{hayami2018microscopic}. 
The odd-parity order parameters are characterized by odd-rank E (MT) multipoles and even-rank ET (M) multipoles in the presence (absence) of the time-reversal symmetry.

\begin{table}[t!]
\caption{
Classification of odd-parity multipoles with respect to the irreducible representation (irrep) of the point group $O_{\rm h}$. 
The superscripts $+$ and $-$ denote time-reversal even and odd, respectively. 
The odd-parity multipoles are shown with their type, rank, notation, and space group symmetry. 
$\nu_{\rm BO}$ and $\nu_{\rm SCO}$ represent the number of modes in each irrep for the bond and spin-current ordered states, respectively. 
}
\label{tab:Odd-parity}
\centering
\renewcommand{\arraystretch}{1.2}
 \begin{tabular}{ccccccc}
 \hline 
 \ \ \ \ irrep. \ \ \ \ & \ \ \ \ type \ \ \ \ & rank & \ notation \ & \ \ symmetry \ \ & 
$\nu_{\rm BO}$ & $\nu_{\rm SCO}$  \\ \hline \hline
A$_{1u}^+$; A$_{1u}^-$ & ET; M &0&$G_0$; $M_0$ &  F4$_1$32 ($O$) & 0; 0 & 1; 1 \\ \hline
A$_{2u}^+$; A$_{2u}^-$ & E; MT &3&$Q_{xyz}$; $T_{xyz}$ & F$\bar{4}$3m ($T_{\rm d}$) & 1; 0 & 1; 0  \\ \hline
E$_{u}^+$; E$_u^-$   & ET; M &2&$G_u$; $M_u$ & I4$_1$22 ($D_4$) & 1; 0 & 2; 1 \\
              & &&$G_v$; $M_v$ & I$\bar{4}$m2 ($D_{2{\rm d}}$) & &  \\ \hline
T$_{1u}^+$; T$_{1u}^-$ & E; MT &1 & $Q_z$; $T_z$& I4$_1$ ($C_4$) & 1; 1 & 2; 2 \\
              & & & $Q_x$; $T_x$ &  &  &  \\
              & & & $Q_y$; $T_y$ &  &  & \\ \hline
T$_{2u}^+$; T$_{2u}^-$ & ET; M & 2 &$G_{xy}$; $M_{xy}$& I$\bar{4}$ ($S_4$) & 0; 1 & 2; 3  \\
              & & & $G_{yz}$; $M_{yz}$ &  &  &\\
              & & & $G_{zx}$; $M_{zx}$ &  &  & \\
\hline 
\end{tabular}
\end{table}

We classify the odd-parity multipoles in Table~\ref{tab:Odd-parity}, up to the rank $l=3$ with respect to the irreducible representation of the cubic $O_{\rm h}$ group in the phase I. 
We also present the space subgroup symmetry for each odd-parity multipole within the symmetries supported by the XRD results~\cite{yamaura2002low}.
Since the XRD measurements~\cite{yamaura2002low} indicate that the space group symmetries in the phases II and III are I$\bar{4}$m2 and I4$_1$22, respectively, and  since Cd$_2$Re$_2$O$_7$ is most likely nonmagnetic (time-reversal even)~\cite{Vyaselev_PhysRevLett.89.017001}, we deduce that the primary order parameters are the ETQs with different components: $G_v$ for the phase II and $G_u$ for the phase III. 
In the following, we examine what types of electronic instability can induce the two ETQs from a microscopic point of view. 

\begin{figure}[t]
\begin{center}
\includegraphics[width=1.0 \hsize]{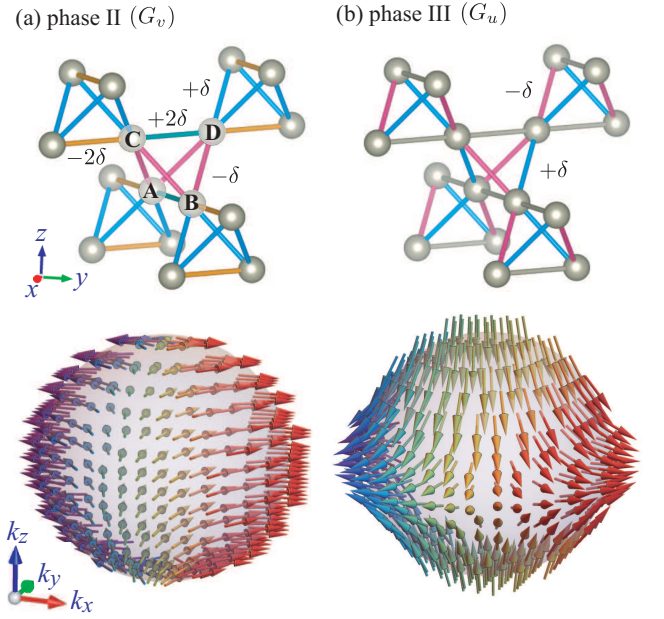} 
\caption{
\label{Fig:bonding}
Schematic pictures of the bond modulations caused by the ETQs: (a) $G_u$ for the phase II and (b) $G_v$ for the phase III. 
Schematic spin polarizations on one of the Fermi surfaces split by the ETQ orderings are shown in the lower figures. 
}
\end{center}
\end{figure}

{\it Electric toroidal quadrupoles.---}
Next, in order to clarify how the ETQs are activated in electronic degrees of freedom, we perform a microscopic analysis based on the tight-binding model. 
While Cd$_2$Re$_2$O$_7$ is a multi-orbital system with relevant $t_{2g}$ orbitals for $5d$ electrons~\cite{Singh_PhysRevB.65.155109,harima2002electronic,Yanagi_comment}, we consider an effective single-orbital model and concentrate on the geometrical effect from the pyrochlore structure composed of the tetrahedron unit.
An extension to multi-orbital models is straightforward by supplementing additional symmetry operations for atomic orbitals at each site. 
The Hamiltonian for the effective tight-binding model is given as 
\begin{align}
\label{eq:Ham_general}
\mathcal{H}=\sum_{\bm{k}\alpha\beta\gamma}\sum_{ij \sigma \sigma'}   c^{\dagger}_{\bm{k}i\sigma} [ \{f^{\rm S}_{\alpha\beta\gamma}(\bm{k}) + f^{\rm A}_{\alpha\beta\gamma}(\bm{k})\}  \rho_\alpha \tau_\beta \sigma_\gamma]^{\sigma\sigma'}_{ij}  c_{\bm{k}j\sigma'}, 
\end{align}
where $c^{\dagger}_{\bm{k}i\sigma}$ ($c_{\bm{k}i\sigma}$) is the creation (annihilation) operator for wave vector $\bm{k}$, sublattice $i=$A-D, and spin $\sigma$. 
Here, the positions of the four sublattice sites within the tetrahedral unit cell are defined by $\bm{r}_{\rm A} =(0, 0, 0)$, $\bm{r}_{\rm B} =(a/4, a/4, 0)$, $\bm{r}_{\rm C} =(a/4, 0, a/4)$, and $\bm{r}_{\rm D} =(0, a/4, a/4)$ [see Fig.~\ref{Fig:bonding}(a); we set $a=1$ as the unit of length]. 
The four sublattice degree of freedom is described by the product of two Pauli matrices $\rho_\alpha$ and $\tau_\beta$; $\rho_\alpha$ spans A-B and C-D, and $\tau_\beta$ spans (AB)-(CD). 
$\sigma_\gamma$ describes $2\times 2$ spin space. 
$f^{\rm S}_{\alpha\beta\gamma}(\bm{k})$ and $f^{\rm A}_{\alpha\beta\gamma}(\bm{k})$
are symmetric and asymmetric form factors with respect to $\bm{k}$, which are related with even and odd-parity multipoles, respectively. 
Note that Eq.~(\ref{eq:Ham_general}) includes all possibilities of 
symmetry-breaking mean fields. 

As the Hamiltonian in Eq.~(\ref{eq:Ham_general}) is an $8\times8$ matrix denoted by the direct product of three Pauli matrices, $\rho_\alpha\tau_\beta\sigma_\gamma$, the total number of independent electronic degrees of freedom is $8 \times 8 \times 2=128$, where the factor $2$ comes from symmetric or antisymmetric nature with respect to $\bm{k}$, i.e., $f^{\rm S}_{\alpha\beta\gamma}$ and $f^{\rm A}_{\alpha\beta\gamma}$.
The 128 electronic degrees of freedom are categorized into the 16 onsite potential types, 96 nearest-neighbor (NN) bond types, and 16 third neighbor bond types. 
Among them, we neglect the 16 onsite potential-type order parameters, as they do not break spatial inversion symmetry. 
We also exclude the 16 third-neighbor bond types because their amplitudes are usually smaller than the NN ones. 
For the remaining 96 NN bond types, we try to elucidate how they activate the ETQs. 

Let us first consider the ETQs without spin degree of freedom. 
In the spinless subspace, the number of electronic degrees of freedom about the NN bond type is reduced to 24. 
They are decomposed into the irreducible representations  $({\rm A}^+_{1g}\oplus {\rm E}^+_{g} \oplus {\rm T}^+_{2g})\oplus ({\rm A}^+_{2u}\oplus {\rm E}^+_{u} \oplus {\rm T}^+_{1u}) \oplus ({\rm T}^-_{1g} \oplus {\rm T}^-_{2g}) \oplus ({\rm T}^-_{1u} \oplus {\rm T}^-_{2u})$, where the superscripts $+$ and $-$ represent time-reversal even and odd, respectively. 
From the decomposition, we find that six types of NN bond modulations can induce odd-parity multipoles of time-reversal even: 
EO $Q_{xyz}$, two ETQs $(G_u, G_v)$, and three E dipoles $(Q_x, Q_y, Q_z)$ (see Table~\ref{tab:Odd-parity}). 
This indicates that the ETQs $G_u$ and $G_v$, which we identified as the order parameters in Cd$_2$Re$_2$O$_7$, can be activated through spontaneous bond orderings (BO). 
By taking an appropriate linear combination of $f^{\rm A}_{\alpha\beta\gamma}(\bm{k})  \rho_\alpha \tau_\beta \sigma_\gamma$~\cite{SM_deri_ETQ}, we obtain the microscopic expressions for the ETQs as 
\begin{eqnarray}
\label{eq:Gu}
G_u &=& \delta [
-s_x c_z \rho_z \tau_y + s_y c_z \rho_y \tau_x + s_z (c_y\rho_x \tau_y - c_x \tau_y) 
 ], \\
 \label{eq:Gv}
G_v &=& \delta [
s_x (2 c_y \rho_y \tau_z -c_z \rho_z \tau_y ) 
+ s_y (2 c_x \rho_y - c_z \rho_y \tau_x) \nonumber \\
& &- s_z (c_x \tau_y + c_y \rho_x \tau_y)
],
\end{eqnarray}
where $s_\mu=\sin (k_\mu/4)$ and $c_\mu=\cos (k_\mu/4)$ ($\mu=x, y, z$), and $\delta$ represents the degree of bond distortions, which corresponds to the order parameter amplitude. 
The bond modulations are schematically shown in Fig.~\ref{Fig:bonding}: (a) for $G_v$ in the phase II [Eq.~(\ref{eq:Gv})] and (b) for $G_u$ in the phase III [Eq.~(\ref{eq:Gu})]. 
Note that each atomic site is no longer the inversion center in these states, reflecting the odd parity of $G_u$ and $G_v$. 
We list the number of modes (independent order parameters) in the BO states in Table~\ref{tab:Odd-parity}.  

We next discuss the ETQs with spin degree of freedom. 
As spins are time-reversal odd, the odd-parity multipoles of time-reversal even are constructed by combining the Pauli matrix $\sigma_\gamma$ and the above spinless odd-parity multipoles with time-reversal odd, i.e., the MT dipoles (MTD) 
$(T_x, T_y, T_z)$ belonging to ${\rm T}^-_{1u}$ and MQs $(M_{yz}, M_{zx}, M_{xy})$ belonging to ${\rm T}^-_{2u}$, as shown in Table~\ref{tab:Odd-parity}.  
By regarding $\sigma_\gamma$ as an axial M dipole belonging to ${\rm T}^-_{2g}$, the odd-parity multipoles of time-reversal even in the spinful case are obtained in the irreducible representations of 
$({\rm T}^-_{1u} \oplus {\rm T}^-_{2u})\otimes {\rm T}^{-}_{2g} \to {\rm A}^+_{1u} \oplus {\rm A}^+_{2u} \oplus 2 {\rm E}^+_u \oplus 2 {\rm T}^+_{1u} \oplus 2{\rm T}^+_{2u}$.  
Consequently, we find four types of active ETQs, $2{\rm E}_{u}^{+}$, whose microscopic expressions are represented by 
\begin{eqnarray}
\label{eq:Gu_s1}
G^{\sigma(1)}_u &=& 2 \sigma_z T_{z} - \sigma_x T_{x} - \sigma_y T_{y}, \\
\label{eq:Gu_s2}
G^{\sigma(2)}_u &=& \sigma_x M_{yz} - \sigma_y M_{zx}, \\ 
\label{eq:Gv_s1}
G^{\sigma(1)}_v &=& \sigma_x T_x - \sigma_y T_y, \\
\label{eq:Gv_s2}
G^{\sigma(2)}_v &=& 2 \sigma_z M_{xy} - \sigma_x M_{yz} - \sigma_y M_{zx}, 
\end{eqnarray}
where the superscript $\sigma$ denotes that the ETQs have spin dependence; 
$(T_x, T_y, T_z)$ and $(M_{yz}, M_{zx}, M_{xy})$ are given as~\cite{SM_deri_ETQ}  
\begin{eqnarray}
\label{eq:Tx}
T_x &=& \delta[s_x (c_y \rho_x + c_z \tau_x)+s_y c_x \rho_x \tau_z + s_z c_x \rho_z \tau_x], \\
\label{eq:Ty}
T_y &=& \delta [s_x c_y \rho_x \tau_z + s_y (c_x \rho_x + c_z \rho_x \tau_x)-s_z c_y \rho_y \tau_y], \\
\label{eq:Tz}
T_z &=& \delta[s_x c_z \rho_z \tau_x -s_y c_z \rho_y \tau_y + s_z (c_x \tau_x + c_y \rho_x \tau_x)], \\
\label{eq:Myz}
M_{yz} &=& \delta[s_x (c_z \tau_x-c_y \rho_x)-s_y c_x \rho_x \tau_z + s_z c_x \rho_z \tau_x], \\
\label{eq:Mzx}
M_{zx} &=& \delta[s_x c_y \rho_x \tau_z + s_y (c_x \rho_x -c_z \rho_x \tau_x)+s_z c_y \rho_y \tau_y], \\
\label{eq:Mxy}
M_{xy} &=& -\delta[s_x c_z \rho_z \tau_x +s_y c_z \rho_y \tau_y - s_z (c_y \rho_x \tau_x -c_x \tau_x)]. 
\end{eqnarray}

\begin{figure}[t]
\begin{center}
\includegraphics[width=1.0 \hsize]{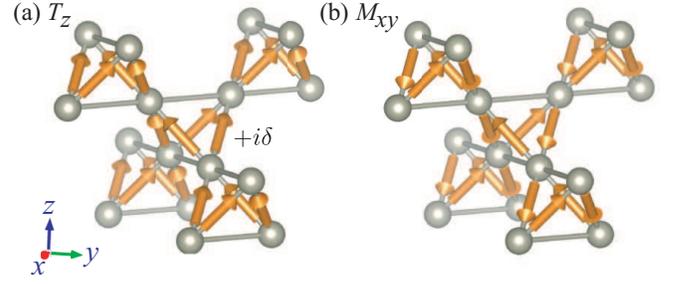} 
\caption{
\label{Fig:bonding_wTRSB}
Schematic pictures of the bond modulations in the presence of (a) the MTD $T_z$ in Eq.~(\ref{eq:Tz}) and (b) the MQ $M_{xy}$ in Eq.~(\ref{eq:Mxy}). 
The arrows on the bonds represent the imaginary hoppings. 
}
\end{center}
\end{figure}

The ETQs in Eqs.~(\ref{eq:Gu_s1})-(\ref{eq:Gv_s2}) are also activated through a bond-order type instability as those in the spinless case in Eqs.~(\ref{eq:Gu}) and (\ref{eq:Gv}). 
However, they originate from asymmetric modulations of time-reversal-odd imaginary hoppings in the spin-dependent form, which can be regarded as spin-current orders (SCOs).  
In Fig.~\ref{Fig:bonding_wTRSB}, we exemplify the MTD $T_z$ [Eq.~(\ref{eq:Tz})] and MQ  $M_{xy}$ [Eq.~(\ref{eq:Mxy})], in which the arrows on each bond represent the imaginary hoppings~\footnote{Note that the imaginary hoppings correspond to the MTD on the bond centers. 
In fact, the BO state in Fig.~\ref{Fig:bonding_wTRSB}(a) has a net MT dipole moment.}.  
This type of SCO has been studied in the context of spontaneous topological Mott insulators~\cite{Raghu_PhysRevLett.100.156401,kurita2011topological}. 
We list the number of modes in the SCO state in Table~\ref{tab:Odd-parity}.

{\it Secondary order parameters.---}
As direct observation of ETQs is rather difficult, we discuss what types of multipoles are additionally induced as the secondary order parameters under the ETQ orders from symmetry arguments~\cite{Hayami_PhysRevB.98.165110}. 
In the phase II with I$\bar{4}$m2 (D$_{2d}$) symmetry, since A$_{2u}$ reduces to symmetric representations A$_1$, the odd-parity EO $Q_{xyz}$ is induced as a secondary order parameter. 
Meanwhile, in the phase III, the odd-parity ET monopole (ETM) $G_{0}$ (time-reversal even pseudoscalar) is induced as a secondary order parameter, since A$_{1u}$ reduces to symmetric representations under the I4$_1$22 (D$_4$) symmetry. 
Furthermore, in both phases II and III, since E$_g$ reduces to symmetric representations A$_1$ $\oplus$ B$_1$, the even-parity E quadrupole (EQ) $Q_{u}$ is induced as a secondary order parameter. 
The observation of these secondary order parameters can be indirect evidences of the ETQ orders. 
For instance, ultrasound and magnetic torque measurements may detect the EQ. 

{\it ETQ-driven phenomena.---}
\begin{table}[t!]
\caption{
Nonzero components of the MC tensor $\alpha_{ij}$ and NRT tensor $\sigma_{ijkl}$ expected under the ETQ orderings. 
The results for the NRT are shown for both primary (op1) and secondary (op2) order parameters. 
See the text for details. 
}
\label{tab:CrossCoupling}
\centering
\begingroup
\renewcommand{\arraystretch}{1.2}
 \begin{tabular}{cl}
\hline
\multicolumn{2}{l}{Phase II  \ \ \ op1: $G_v$ (ETQ), op2: $Q_{xyz}$ (EO), $Q_u$ (EQ)}  \\ \hline \hline
\ \ MC \ \ \ & $\alpha_{xx}=-\alpha_{yy}$  \\ \hline
\ \ NRT 
 \ \ \ & $ \sigma_{xxxx}=-\sigma_{yyyy}$, $\sigma_{xxyy}=-\sigma_{yyxx}$, $\sigma_{xyyx}=-\sigma_{yxxy}$, \\
& $\sigma_{xxzz}=-\sigma_{yyzz}$, $\sigma_{xzzx}=-\sigma_{yzzy}$, $\sigma_{zzxx}=-\sigma_{zzyy}$, \\
& $\sigma_{zxxz}=-\sigma_{zyyz}$  \\
\hline
\multicolumn{2}{l}{Phase III \ \ op1: $G_u$ (ETQ), op2: $G_0$ (ETM), $Q_u$ (EQ)}
\\ \hline \hline
\ \ MC \ \ \ & $\alpha_{xx}=\alpha_{yy}$, \ \ $\alpha_{zz}$ \\ \hline
\ \ NRT 
 \ \ \ & $\sigma_{xxxx}=\sigma_{yyyy}$, \ \ $\sigma_{xxyy}=\sigma_{yyxx}$, \ \ $\sigma_{xyyx}=\sigma_{yxxy}$, \\
& $\sigma_{xxzz}=\sigma_{yyzz}$, \ \ $\sigma_{xzzx}=\sigma_{yzzy}$, \ \ $\sigma_{zzxx}=\sigma_{zzyy}$, \\
& $\sigma_{zxxz}=\sigma_{zyyz}$, \ \ $\sigma_{zzzz}$ \\
\hline 
\end{tabular}
\endgroup
\end{table}
For further identification of the ETQs, we discuss physical phenomena driven by the ETQ orderings. 
As the ETQs break spatial inversion symmetry, the band structures in both phases II and III exhibit spin splitting as the Rashba metals~\cite{Fu_PhysRevLett.115.026401,harter2017parity}. 
The origin of such spin splitting is the ASOC induced by the ETQ orderings. 
The functional form of the ASOC is derived by considering the active odd-parity multipoles belonging to symmetric representations in their space groups~\cite{Hayami_PhysRevB.98.165110,watanabe2018group}; namely, $G_v$ and $Q_{xyz}$ in the phase II, and $G_u$ and $G_0$ in the phase III. 
In particular, the odd-parity multipoles with rank 0-2 lead to the ASOC in the first order of $\bm{k}$~\cite{Hayami_PhysRevB.98.165110}. 
The resultant functional forms of the ASOC for the phases II and III are given by 
\begin{eqnarray}
\mathcal{H}^{\rm II}_{\rm ASOC} &=& c_1 (k_x \sigma_x - k_y \sigma_y)+\mathcal{O}(k^3), \\
\mathcal{H}^{\rm III}_{\rm ASOC} &=& c_1 (k_x \sigma_x + k_y \sigma_y)+c_2 k_z \sigma_z+\mathcal{O}(k^3), 
\end{eqnarray} 
respectively, where $c_1$ and $c_2$ are appropriate constants proportional to the order parameter amplitude $\delta$. 
The spin polarizations on the spin-split Fermi surface are schematically shown for the phases II and III in the lower pictures of Figs.~\ref{Fig:bonding}(a) and \ref{Fig:bonding}(b), respectively. 
Note that such spin splitting in the band structure occurs even for the spinless ETQs in Eqs.~(\ref{eq:Gu}) and (\ref{eq:Gv}) in the presence of the spin-orbit coupling. 

In addition, the ETQs give rise to intriguing responses to external stimuli. 
One is the MC effect, in which a uniform magnetization $M_i$ is induced by an electric current $J_j$ ($i,j=x,y,z$) as~\cite{Hayami_PhysRevB.98.165110,watanabe2018group} 
\begin{eqnarray}
M_i=\alpha_{ij} J_j, 
\end{eqnarray}
where the MC tensor $\alpha_{ij}$ is the rank-2 axial tensor of time-reversal even. 
The form of $\alpha_{ij}$ is related with active odd-parity multipoles with rank 0-2 $(G_0\oplus Q_{1m} \oplus G_{2m})$~\cite{Hayami_PhysRevB.98.165110}; the rank-0, 1, and 2 multipoles have the isotropic, antisymmetric, and symmetric traceless components, respectively.  
Thus, $\alpha_{ij}$ in the phase II becomes symmetric and traceless corresponding to $G_{v}$: $\alpha_{xx}=-\alpha_{yy}\propto G_{v}$. 
Meanwhile, $\alpha_{ij}$ in the phase III has two nonzero symmetric components reflecting $G_{0}$ and $G_u$: $\alpha_{xx}=\alpha_{yy}\propto G_{0}-G_{u}$ 
 and $\alpha_{zz}\propto G_{0}+2G_{u}$. 
The results are summarized in Table~\ref{tab:CrossCoupling}. 

Another interesting response is the NRT. 
As a nonreciprocal current is proportional to the second order of an electric field, the NRT needs the breaking of time-reversal symmetry by an external magnetic field as 
\begin{eqnarray}
\label{eq:NRD}
J_i=\sigma_{ijkl} E_j E_k H_l, 
\end{eqnarray}
where the NRT tensor $\sigma_{ijkl}$ is the rank-4 axial tensor; $E_j$ and $H_l$ are electric and magnetic fields, respectively. 
From symmetry arguments, the form of the NRT tensor is related with the multipoles with rank 0-4 $(2G_0\oplus 3Q_{1m} \oplus 4G_{2m} \oplus 2Q_{3m} \oplus G_{4m})$~\cite{Hayami_PhysRevB.98.165110}. 
Consequently, $\sigma_{ijkl}$ has independent seven (eight) components in the phase II (III), as shown in Table~\ref{tab:CrossCoupling}. 
We note that higher-order ET hexadecapoles also become active: $G_{4v}$ in the phase II, and $G_4$ and $G_{4u}$ in the phase III. 

It is also interesting to point out that a lattice distortion is induced by an electric current in an applied magnetic field as $\zeta_{ij}=d_{ijkl} J_k H_l $ where $d_{ijkl}$ represents a strain tensor~\cite{Hayami_PhysRevB.98.165110,watanabe2018group}. 
This is easily understood by noting that $E_j E_k$ in Eq.~(\ref{eq:NRD}) and $\zeta_{ij}$ show the same transformation under the space-time inversion. 
Thus, the tensor $d_{ijkl}$ has similar nonzero components as $\sigma_{ijkl}$ in Eq.~(\ref{eq:NRD}).

{\it Conclusion.---}
We theoretically showed that the odd-parity ETQs with different components of $x^2-y^2$ and $3z^2-r^2$ are the candidates of the primary order parameters in the phases II and III, respectively, in the spin-orbit coupled metal Cd$_2$Re$_2$O$_7$. 
We clarified that electronic instabilities toward spontaneous bond or spin-current ordering on Re-Re bonds induce the ETQ orders. 
We also discussed how to identify the ETQs by exemplifying their characteristic phenomena, such as spin-split Fermi surfaces, MC effect, and NRT in an applied magnetic field. 
Our ETQ scenario will give an insight into the origin of the enigmatic phase transitions in Cd$_2$Re$_2$O$_7$. 
Furthermore, our microscopic classifications of multipoles are widely applicable to other spontaneously parity-breaking systems.

\begin{acknowledgments}
The authors thank Z. Hiroi, J. Yamaura, S. Uji, D. Hirai, and H. Hirose for the fruitful discussions on experimental information in Cd$_2$Re$_2$O$_7$.
This research was supported by JSPS KAKENHI Grants Numbers JP15H05885, JP18H04296 (J-Physics), and JP18K13488. 
\end{acknowledgments}

\bibliographystyle{apsrev}
\bibliography{ref}

\newpage
\begin{widetext}
\appendix

\section{Supplemental material for \\
``Electric Toroidal Quadrupoles in Spin-Orbit Coupled Metal Cd$_2$Re$_2$O$_7$''}

\subsection{Microscopic expressions of electric toroidal quadrupoles}
We present a derivation of the microscopic expressions of ETQs in Eqs.~(2), (3), (8)-(13) in the main text. 
First, we consider multipole degrees of freedom on a unit of tetrahedron. 
The irreducible representation of molecular orbitals under the $T_{\rm d}$ group is given by ${\rm A}_1 \oplus {\rm T}_{2}$. 
Then, the basis wave functions are expressed as 
\begin{align}
{\rm A}_{1}:& \psi_{{\rm A}_1} = \frac{1}{2} (\psi_{\rm A} + \psi_{\rm B} + \psi_{\rm C} + \psi_{\rm D}), \\
{\rm T}_{2}:& \psi_{{\rm T}_{2,x}} = \frac{1}{2} (-\psi_{\rm A} + \psi_{\rm B} + \psi_{\rm C} - \psi_{\rm D}), \\
& \psi_{{\rm T}_{2,y}} = \frac{1}{2} (-\psi_{\rm A} + \psi_{\rm B} - \psi_{\rm C} + \psi_{\rm D}), \\
& \psi_{{\rm T}_{2,z}} = \frac{1}{2} (-\psi_{\rm A} - \psi_{\rm B} + \psi_{\rm C} + \psi_{\rm D}), 
\end{align}
where $\psi_{i}$ is the $s$-wave-like atomic wave function at site $i=$A-D [see Fig.~1(a) in the main text]. 
For these basis functions, there are sixteen electronic degrees of freedom, which are decomposed by the irreducible representation of the symmetry of a unit of tetrahedron, $T_{\rm d}$, as 
\begin{align}
({\rm A}_1 \oplus {\rm T}_{2})\otimes ({\rm A}_1 \oplus {\rm T}_{2})
= (2{\rm A}^+_{1} \oplus {\rm E}^+ \oplus 2{\rm T}^{+}_{2}) \oplus ({\rm T}^{-}_{1}\oplus {\rm T}^{-}_{2}).  
\end{align} 
As the molecular orbitals $\psi_{{\rm A}_1}$ and $\psi_{{\rm T}_{2}}=(\psi_{{\rm T}_{2,x}}, \psi_{{\rm T}_{2,y}}, \psi_{{\rm T}_{2,z}})$ have the same symmetry properties as the atomic orbitals $s$ and $(p_{x}, p_{y}, p_{z})$, respectively, these electronic degrees of freedom are represented by sixteen independent multipoles in an $s$-$p$ hybridized system~\cite{hayami2018microscopic}: two electric monopoles $\tilde{Q}^{(1)}_0$ and $\tilde{Q}^{(2)}_0$ belonging to ${\rm A}^{+}_{1}$, two electric quadrupoles $(\tilde{Q}_u, \tilde{Q}_v)$ belonging to ${\rm E}^+$, three electric dipoles $(\tilde{Q}_{x}, \tilde{Q}_{y}, \tilde{Q}_{z})$ and other three electric quadrupoles $(\tilde{Q}^{(1)}_{yz}, \tilde{Q}^{(1)}_{zx}, \tilde{Q}^{(1)}_{xy})$  belonging to ${\rm T}^+_{2}$, three magnetic dipoles $(\tilde{M}_x, \tilde{M}_y, \tilde{M}_z)$ belonging to ${\rm T}^-_{1}$, and three magnetic toroidal dipoles $(\tilde{T}_{x}, \tilde{T}_{y}, \tilde{T}_{z})$ belonging to ${\rm T}^-_{2}$. 
The matrix elements for each multipole in the basis functions $(\psi_{{\rm A}_1}, \psi_{{\rm T}_{2,x}}, \psi_{{\rm T}_{2,y}}, \psi_{{\rm T}_{2,z}})$ are given by~\cite{hayami2018microscopic}  
\begin{align}
&
\tilde{Q}^{(1)}_{0}=
\left(
\begin{array}{cccc}
 1 & 0 & 0 & 0 \\ 
 0 & 1 & 0 & 0 \\
 0 & 0 & 1 & 0 \\
 0 & 0 & 0 & 1 \\
\end{array}
\right),
\ \ 
\tilde{Q}^{(2)}_{0}=
\left(
\begin{array}{cccc}
 3 & 0 & 0 & 0 \\ 
 0 & -1 & 0 & 0 \\
 0 & 0 & -1 & 0 \\
 0 & 0 & 0 & -1 \\
\end{array}
\right),
\\ &
\label{eq:ED}
\tilde{Q}_{x}=
\left(
\begin{array}{cccc}
 0 & -2 & 0 & 0 \\ 
 -2 & 0 & 0 & 0 \\
 0 & 0 & 0 & 0 \\
 0 & 0 & 0 & 0 \\
\end{array}
\right),
\ \
\tilde{Q}_{y}=
\left(
\begin{array}{cccc}
 0 & 0 & -2 & 0 \\ 
 0 & 0 & 0 & 0 \\
 -2 & 0 & 0 & 0 \\
 0 & 0 & 0 & 0 \\
\end{array}
\right),
\ \
\tilde{Q}_{z}=
\left(
\begin{array}{cccc}
 0 & 0 & 0 & -2 \\ 
 0 & 0 & 0 & 0 \\
 0 & 0 & 0 & 0 \\
 -2 & 0 & 0 & 0 \\
\end{array}
\right),
\\ &
\tilde{Q}_{u}=
\left(
\begin{array}{cccc}
 0 & 0 & 0 & 0 \\ 
 0 & 2 & 0 & 0 \\
 0 & 0 & 2 & 0 \\
 0 & 0 & 0 & -4 \\
\end{array}
\right),
\ \
\tilde{Q}_{v}=
\left(
\begin{array}{cccc}
 0 & 0 & 0 & 0 \\ 
 0 & -2 & 0 & 0 \\
 0 & 0 & 2 & 0 \\
 0 & 0 & 0 & 0 \\
\end{array}
\right),
\\&
\tilde{Q}^{(1)}_{yz}=
\left(
\begin{array}{cccc}
 0 & 0 & 0 & 0 \\ 
 0 & 0 & 0 & 0 \\
 0 & 0 & 0 & 2 \\
 0 & 0 & 2 & 0 \\
\end{array}
\right),
\ \ 
\tilde{Q}^{(1)}_{zx}=
\left(
\begin{array}{cccc}
 0 & 0 & 0 & 0 \\ 
 0 & 0 & 0 & 2 \\
 0 & 0 & 0 & 0 \\
 0 & 2 & 0 & 0 \\
\end{array}
\right),
\ \ 
\tilde{Q}^{(1)}_{xy}=
\left(
\begin{array}{cccc}
 0 & 0 & 0 & 0 \\ 
 0 & 0 & 2 & 0 \\
 0 & 2 & 0 & 0 \\
 0 & 0 & 0 & 0 \\
\end{array}
\right), 
\\&
\tilde{M}_{x}=
\left(
\begin{array}{cccc}
 0 & 0 & 0 & 0 \\ 
 0 & 0 & 0 & 0 \\
 0 & 0 & 0 & -2i \\
 0 & 0 & 2i & 0 \\
\end{array}
\right),
\ \
\tilde{M}_{y}=
\left(
\begin{array}{cccc}
 0 & 0 & 0 & 0 \\ 
 0 & 0 & 0 & 2i \\
 0 & 0 & 0 & 0 \\
 0 & -2i & 0 & 0 \\
\end{array}
\right),
\ \ 
\tilde{M}_{z}=
\left(
\begin{array}{cccc}
 0 & 0 & 0 & 0 \\ 
 0 & 0 & -2i & 0 \\
 0 & 2i & 0 & 0 \\
 0 & 0 & 0 & 0 \\
\end{array}
\right),
\\&
\label{eq:MTD}
\tilde{T}_{x}=
\left(
\begin{array}{cccc}
 0 & -2 i & 0 & 0 \\ 
 2 i & 0 & 0 & 0 \\
 0 & 0 & 0 & 0 \\
 0 & 0 & 0 & 0 \\
\end{array}
\right),
\ \ 
\tilde{T}_{y}=
\left(
\begin{array}{cccc}
 0 & 0 & -2 i & 0 \\ 
 0 & 0 & 0 & 0 \\
 2 i & 0 & 0 & 0 \\
 0 & 0 & 0 & 0 \\
\end{array}
\right),
\ \ 
\tilde{T}_{z}=
\left(
\begin{array}{cccc}
 0 & 0 & 0 & -2 i \\ 
 0 & 0 & 0 & 0 \\
 0 & 0 & 0 & 0 \\
 2i & 0 & 0 & 0 \\
\end{array}
\right),  
\end{align}
where we take appropriate normalizations to simplify the following expressions. Note that $(\tilde{Q}_x, \tilde{Q}_y, \tilde{Q}_z)$ in Eq.~(\ref{eq:ED}) and $(\tilde{T}_x, \tilde{T}_y, \tilde{T}_z)$ in Eq.~(\ref{eq:MTD}) are the odd-parity multipoles. 
In the following calculations for the centrosymmetric pyrochlore structure (point group ${\rm O}_h$) consisting of upward and downward tetrahedra, these odd-parity multipoles are replaced by the even-parity multipoles in the same irreducible representation as $(\tilde{Q}_x, \tilde{Q}_y, \tilde{Q}_z) \to (\tilde{Q}^{(2)}_{yz}, \tilde{Q}^{(2)}_{zx}, \tilde{Q}^{(2)}_{xy})$ and $(\tilde{T}_x, \tilde{T}_y, \tilde{T}_z) \to (\tilde{T}_{yz}, \tilde{T}_{zx}, \tilde{T}_{xy})$. 
By a unitary transformation, the multipole degrees of freedom for a unit of tetrahedron in the pyrochlore structure in the basis $(\psi_{\rm A}, \psi_{\rm B}, \psi_{\rm C}, \psi_{\rm D})$ can be expressed as 
\begin{align}
\label{eq:A1g}
&{\rm A}^+_{1g}: Q^{(1)}_0=1, Q^{(2)}_0 = \rho_x + \tau_x + \rho_x \tau_x,  \\
&{\rm E}^+_{g}: (Q_u, Q_v)=(\tau_x - 2 \rho_x + \rho_x \tau_x, \tau_x -\rho_x \tau_x), \\
&{\rm T}^+_{2g}: (Q^{(1)}_{yz}, Q^{(1)}_{zx}, Q^{(1)}_{xy})=(\rho_z \tau_z + \rho_y \tau_y, \rho_z -\rho_z \tau_x, \tau_z - \rho_x \tau_z), \\
&{\rm T}^+_{2g}: (Q^{(2)}_{yz}, Q^{(2)}_{zx}, Q^{(2)}_{xy})=(\rho_z \tau_z - \rho_y \tau_y, \rho_z +\rho_z \tau_x, \tau_z + \rho_x \tau_z), \\
&{\rm T}^-_{1g}: (M_{x}, M_{y}, M_{z})=(\rho_y \tau_z - \rho_z \tau_y, -\rho_y +\rho_y \tau_x, \tau_y - \rho_x \tau_y), \\
\label{eq:T2g}
&{\rm T}^-_{2g}: (T_{yz}, T_{zx}, T_{xy})=(\rho_y \tau_z + \rho_z \tau_y, \rho_y +\rho_y \tau_x, \tau_y + \rho_x \tau_y),   
\end{align}
where $\rho_\alpha$ and $\tau_\beta$ are $2\times 2$ Pauli matrices representing physical spaces spanned by A-B and C-D, and (AB)-(CD), respectively (see the main text). 

Next, we take into account bond degrees of freedom by connecting upward and downward tetrahedra in the pyrochlore structure. 
As the multipole degrees of freedom in Eqs.~(\ref{eq:A1g})-(\ref{eq:T2g}) are even-parity, asymmetric bond modulations with respect to each atomic site are necessary for describing spontaneous parity-breaking orders. 
By focusing on the nearest-neighbor bond modulations (see the main text), we can introduce a mean-field Hamiltonian for spontaneous parity breaking in the following form: 
\begin{align}
\mathcal{H}^{\rm op}(\bm{k})&=
\left(
\begin{array}{cccc}
 0 & -i \delta_{\rm AB} \sin \left(\frac{k_x}{4} + \frac{k_y}{4}\right) & -i \delta_{\rm AC} \sin\left(\frac{k_x}{4} + \frac{k_z}{4}\right) & -i \delta_{\rm AD}\sin\left(\frac{k_y}{4} + \frac{k_z}{4}\right) \\ 
 i \delta^*_{\rm AB} \sin \left(\frac{k_x}{4} + \frac{k_y}{4}\right) & 0 & i \delta_{\rm BC} \sin\left(\frac{k_y}{4} -\frac{k_z}{4}\right) & i \delta_{\rm BD} \sin\left(\frac{k_x}{4} -\frac{k_z}{4}\right) \\
 i \delta^*_{\rm AC} \sin \left(\frac{k_x}{4} + \frac{k_z}{4}\right) & -i \delta^*_{\rm BC} \sin\left(\frac{k_y}{4}-\frac{k_z}{4}\right) & 0 & i \delta_{\rm CD} \sin\left(\frac{k_x}{4} - \frac{k_y}{4}\right) \\
 i \delta^*_{\rm AD} \sin\left(\frac{k_y}{4} + \frac{k_z}{4}\right) & -i \delta^*_{\rm BD} \sin\left(\frac{k_x}{4}-\frac{k_z}{4}\right) & -i \delta^*_{\rm CD}\sin\left(\frac{k_x}{4} -\frac{k_y}{4}\right) & 0 \\
\end{array}
\right) \\
&=\frac{1}{2} (\rho_y+\rho_y\tau_z){\rm Re}[\delta_{\rm AB}]\sin\left(\frac{k_x}{4} + \frac{k_y}{4}\right)-\frac{1}{2}  (\rho_y-\rho_y\tau_z){\rm Re}[\delta_{\rm CD}]\sin\left(\frac{k_x}{4} -\frac{k_y}{4} \right) \nonumber \\
&+\frac{1}{2} (\tau_y+\rho_z \tau_y) {\rm Re}[\delta_{\rm AC}]\sin\left(\frac{k_x}{4} + \frac{k_z}{4}\right) 
-\frac{1}{2}(\tau_y-\rho_z \tau_y) {\rm Re}[\delta_{\rm BD}] \sin\left(\frac{k_x}{4}-\frac{k_z}{4}\right) \nonumber \\
&+\frac{1}{2}(\rho_x \tau_y + \rho_y \tau_x) {\rm Re}[\delta_{\rm AD}]\sin\left(\frac{k_y}{4} + \frac{k_z}{4}\right) -\frac{1}{2}(\rho_x \tau_y - \rho_y \tau_x) {\rm Re}[\delta_{\rm BC}] \sin \left(\frac{k_y}{4} -\frac{k_z}{4}\right) \nonumber \\
&   +\frac{1}{2}(\rho_x+\rho_x\tau_z){\rm Im}[\delta_{\rm AB}]\sin\left(\frac{k_x}{4} + \frac{k_y}{4}\right)- \frac{1}{2} (\rho_x-\rho_x\tau_z){\rm Im}[\delta_{\rm CD}]\sin\left(\frac{k_x}{4} -\frac{k_y}{4} \right) \nonumber \\
&+\frac{1}{2} (\tau_x+\rho_z \tau_x) {\rm Im}[\delta_{\rm AC}]\sin\left(\frac{k_x}{4} + \frac{k_z}{4}\right) 
-\frac{1}{2}(\tau_x-\rho_z \tau_x) {\rm Im}[\delta_{\rm BD}] \sin\left(\frac{k_x}{4}-\frac{k_z}{4}\right) \nonumber \\
&+\frac{1}{2}(\rho_x \tau_x - \rho_y \tau_y) {\rm Im}[\delta_{\rm AD}]\sin\left(\frac{k_y}{4} + \frac{k_z}{4}\right) -\frac{1}{2}(\rho_x \tau_x + \rho_y \tau_y) {\rm Im}[\delta_{\rm BC}] \sin \left(\frac{k_y}{4} -\frac{k_z}{4}\right), 
\label{Eq:bond_OP}
\end{align}
where $\delta_{ij}={\rm Re}[\delta_{ij}]+i{\rm Im}[\delta_{ij}]$ is the complex variable describing the bond modulation between sublattices $i$ and $j$ ($i,j=$A-D); the real and imaginary parts of $\delta_{ij}$ represent the bond modulations with time-reversal even and odd, respectively. 
$\mathcal{H}^{\rm op}(\bm{k})$ is a part of $ [  f^{\rm A}_{\alpha\beta0}(\bm{k})  \rho_\alpha \tau_\beta ]_{ij}^{\sigma\sigma}$ in Eq.~(1) in the main text. 
By writing down the bond degrees of freedom as $\tilde{\delta}=(\delta_{\rm AB}, \delta_{\rm AC}, \delta_{\rm AD}, \delta_{\rm BC}, \delta_{\rm BD}, \delta_{\rm CD})$, the irreducible representations for the bond modulations are obtained as $\tilde{\delta}_{Q_{xyz}}=\delta(1, 1, 1, 1, 1, 1)$ belonging to ${\rm A}^+_{2u}$,  $(\tilde{\delta}_{G_{u}}, \tilde{\delta}_{G_{v}})=[\delta(2, -1, -1, -1, -1, 2), \delta(0,1,-1,-1,1,0)]$ belonging to ${\rm E}^+_u$, and $(\tilde{\delta}_{Q_{x}}, \tilde{\delta}_{Q_{y}}, \tilde{\delta}_{Q_{z}})=[\delta(0, 0, -1, 1, 0, 0), \delta(0,-1,0,0,1,0), \delta(-1,0,0,0,0,1)]$ belonging to ${\rm T}^+_{1u}$ with time-reversal even, while they are given by $(\tilde{\delta}_{T_{x}}, \tilde{\delta}_{T_{y}}, \tilde{\delta}_{T_{z}})=[\delta(i, i, 0, 0, -i, -i), \delta(i,0,i,-i,0,i), \delta(0,i,i,i,i,0)]$ belonging to ${\rm T}^-_{1u}$ and $(\tilde{\delta}_{M_{yz}}, \tilde{\delta}_{M_{zx}}, \tilde{\delta}_{M_{xy}})=[\delta(-i, i, 0, 0, -i, i), \delta(i,0,-i,i,0,i), \delta(0,-i,i,i,-i,0)]$ belonging to ${\rm T}^-_{2u}$ with time-reversal odd. 
By substituting these modulations into Eq.~(\ref{Eq:bond_OP}), the microscopic expressions for ETQs are obtained as Eqs.~(2), (3), (8)-(13) in the main text 
Note that these expressions can be rewritten in terms of the multipole degrees of freedom in Eqs.~(\ref{eq:A1g})-(\ref{eq:T2g}), which are given by 
\begin{align}
G_u=&\frac{1}{2} \left[c_x s_x(M_x-T_{yz}) +c_z s_y (M_y+T_{zx}) - s_z \left\{c_x (M_z + T_{xy})+ c_y (M_z-T_{xy})\right\}\right], \\
G_v=&\frac{1}{2} [ s_x \left\{ 2 c_y (M_x+ T_{yz})+ c_z (M_x - T_{yz}) \right\}
-s_y\left\{2 c_x ( M_y  -T_{zx})+ c_z  (M_y+ T_{zx})\right\} 
\nonumber \\
 &-s_z\left\{ c_x (M_z+ T_{xy}) - c_y (M_z -T_{xy})\right\} ], \\
T_x=&\frac{1}{6} \left[3 c_x \left\{s_y(Q_{xy}^{(2)}-Q_{xy}^{(1)})+s_z(Q_{zx}^{(2)}-Q_{zx}^{(1)} )\right\}
+s_x\left\{c_y (2 Q_0^{(2)}+Q'_u+3 Q'_v)+c_z (2 Q_0^{(2)}+Q'_u-3 Q'_v)\right\}\right], \\
T_y=&\frac{1}{6} \left[3 c_y \left\{s_z(Q_{yz}^{(2)} -Q_{yz}^{(1)})+s_x(Q_{xy}^{(2)}-Q_{xy}^{(1)}) \right\}
+s_y \left\{c_z (2 Q_0^{(2)}+Q''_u+3 Q''_v)+c_x (2 Q_0^{(2)}+Q''_u-3 Q''_v)\right\}\right], \\
T_z=&\frac{1}{6} \left[3 c_z \left\{s_x(Q_{zx}^{(2)}-Q_{zx}^{(1)})+s_y(Q_{yz}^{(2)}-Q_{yz}^{(1)})\right\}
+s_z \left\{c_x (2 Q_0^{(2)}+Q_u+3 Q_v)+c_y (2 Q_0^{(2)}+Q_u-3 Q_v)\right\}\right], \\
M_{yz}=&\frac{1}{6} \left[3 c_x \left\{-s_y(Q_{xy}^{(2)}- Q_{xy}^{(1)})+s_z(Q_{zx}^{(2)}-Q_{zx}^{(1)})\right\}-s_x\left\{c_y (2 Q_0^{(2)}+Q'_u+3 Q'_v)-c_z (2 Q_0^{(2)}+Q'_u-3 Q'_v)\right\}\right], \\
M_{zx}=&\frac{1}{6} \left[3 c_y \left\{-s_z(Q_{yz}^{(2)}-Q_{yz}^{(1)})+s_x(Q_{xy}^{(2)}-Q_{xy}^{(1)}) \right\}-s_y \left\{c_z (2 Q_0^{(2)}+Q''_u+3 Q''_v)-c_x (2 Q_0^{(2)}+Q''_u-3 Q''_v)\right\}\right],\\
M_{xy}=&\frac{1}{6} \left[3 c_z \left\{-s_x(Q_{zx}^{(2)}-Q_{zx}^{(1)})+s_y(Q_{yz}^{(2)}-Q_{yz}^{(1)}) \right\}-s_z \left\{c_x (2 Q_0^{(2)}+Q_u+3 Q_v)-c_y (2 Q_0^{(2)}+Q_u-3 Q_v)\right\}\right],
\end{align}
where we define $2(Q'_u, Q'_v)=(-Q_u+3Q_v,-Q_u-Q_v)\propto (3x^2-r^2, y^2-z^2)$ and $2(Q''_u, Q''_v)=(-Q_u-3Q_v,Q_u-Q_v)\propto (3y^2-r^2, z^2-x^2)$ for notational simplicity. 

\end{widetext}
\end{document}